\documentstyle[epsf,aps]{revtex}
\newcommand{\mafigura}[4]{
  \begin{figure}[hbtp]
    \begin{center}
      \epsfxsize=#1 \leavevmode \epsffile{#2}
    \end{center}
    \caption{#3}
    \label{#4}
  \end{figure} }
\newcommand{\eq}[1]{eq.~(\ref{#1})}
\newcommand{\beq}{\begin{equation}}
\newcommand{\eeq}{\end{equation}}
\newcommand{\bea}{\begin{eqnarray}}
\newcommand{\eea}{\end{eqnarray}}
\newcommand{\non}{\nonumber}
\begin{document}
\preprint{TTP98-15}
\preprint{hep-ph/9807420}

\title{Charge asymmetry of heavy quarks at hadron colliders}

\draft

\author{J.H. K\"uhn}
\address{Institut f\"ur Theoretische Teilchenphysik,
Universit\"at Karlsruhe, D-76128 Karlsruhe, Germany}
\author{G. Rodrigo}
\address{INFN-Sezione di Firenze,
Largo E. Fermi 2, I-50125 Firenze, Italy}

%\date{\today}
\date{July 20, 1998}
 
\maketitle

\begin{abstract}
A sizeable difference in the differential production cross section
of top and antitop quarks, respectively, is predicted for 
hadronically produced heavy quarks.
It is of order $\alpha_s$ and arises from the 
interference between charge odd and even amplitudes respectively.
For the TEVATRON it amounts up to 15\% for the differential 
distribution in suitable chosen kinematical regions.
The resulting integrated forward-backward asymmetry of 
4--5\% could be measured in the next round of experiments.
At the LHC the asymmetry can be studied by selecting 
appropriately chosen kinematical regions.
Furthermore, a slight preference at LHC for centrally 
produced antitop is predicted, with top quarks more abundant at 
large positive and negative rapidities.
\end{abstract}


%\narrowtext

%%%%%%%%%%%%%%%%%%%%%%%%%%%%%%%%%%%%%%%%%%%%%%%%%%%%%%%%%%%%
%%%%%%%%%%%%%%%%%%%%%%%%%%%%%%%%%%%%%%%%%%%%%%%%%%%%%%%%%%%%

\section{Introduction}

Heavy flavor production at hadron colliders is one of the most
active fields of current theoretical and experimental studies.
Large event rates, combined with improved experimental
techniques, allow for detailed investigations of the 
properties of heavy quarks and their production mechanism
at the same time.
While charm production with a quark mass around $1.5$~GeV is 
barely accessible to perturbative QCD calculations, bottom and 
{\it a forteriori} top production should be well described by this 
approach.

Theoretical and experimental
results~\cite{Catani:1997rn,Tipton:1996}
for the cross section of hadronic top production are well
consistent with this expectation.
Obviously, in view of the large QCD coupling, the inclusion 
of higher order QCD corrections in these calculations is 
mandatory for a successful comparison. Recent studies have, 
to a large extent, concentrated on the predictions of the 
total cross section and a few selected one particle 
inclusive distributions.
In this paper a different issue of heavy flavor production 
is investigated, namely the charge asymmetry,
which is sensitive toward
a specific subclass of virtual and real radiative corrections.

Evaluated in Born approximation 
the lowest order processes relevant for heavy flavor 
production
\beq
q + \bar{q} \to Q + \bar{Q}~,
\label{eq:qqbar}
\eeq
\beq
g + g \to Q + \bar{Q}~,
\label{eq:gg}
\eeq
do not discriminate between the final quark and antiquark, thus  
predicting identical differential distributions  also for 
the hadronic production process.
However, radiative corrections involving either virtual or 
real gluon emission lead to a sizeable difference between 
the differential quark and antiquark production process 
and hence to a charge asymmetry which could be well 
accessible experimentally.

This asymmetry has its origin in two different reactions:
radiative corrections to quark-antiquark fusion (Fig.~\ref{fig:qqbar})
and heavy flavor production involving interference terms
of different amplitudes contributing to 
gluon-quark scattering (Fig.~\ref{fig:qg}) 
\beq
g+q \to Q + \bar{Q} + q~,
\label{eq:gq}
\eeq
a reaction intrinsically of order $\alpha_s^3$.
Gluon fusion remains of course charge symmetric.
In both reactions (\ref{eq:qqbar}) and (\ref{eq:gq})
the asymmetry can be traced to the 
interference between amplitudes which are relatively odd under the 
exchange of $Q$ and $\bar{Q}$. 
In fact, as shown below in detail, the asymmetry can be understood
in analogy to the corresponding one in QED reactions
and is proportional to the color factor $d_{abc}^2$.
In contrast, the non-Abelian contributions,
in particular those involving the triple gluon coupling, 
lead to symmetric pieces in the 
differential cross section.
Event generators which do not include the full next-to-leading matrix
elements~\cite{Marchesini:1991ch,Sjostrand:1994yb} cannot 
predict the asymmetry.

%%%%%%%%%%%%%%%
\mafigura{8 cm}{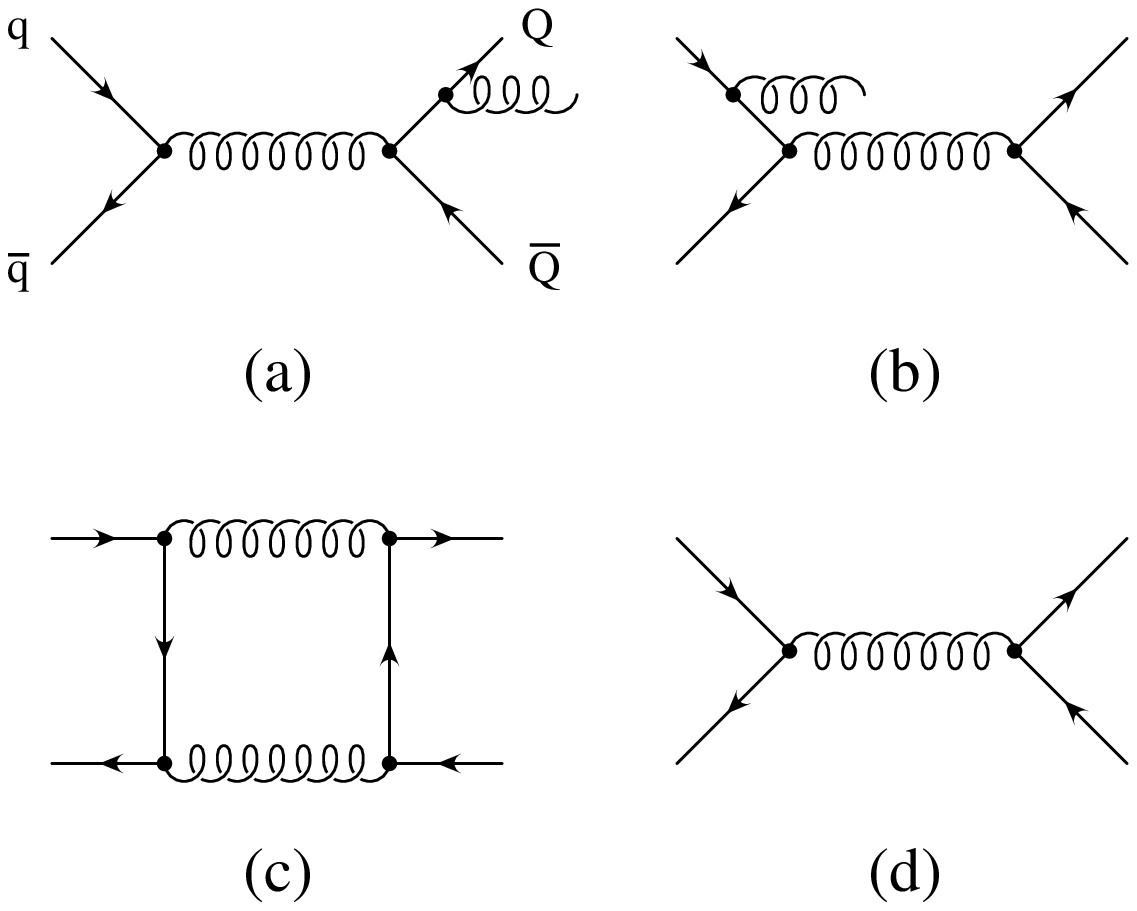}{Origin of the QCD charge
asymmetry in hadroproduction of heavy quarks:
interference of final-state (a) with initial-state (b) gluon bremsstrahlung
plus interference of the box (c) with the Born diagram (d).
Only representative diagrams are shown.}
{fig:qqbar}
%%%%%%%%%%%%%%%
%%%%%%%%%%%%%%%
\mafigura{12 cm}{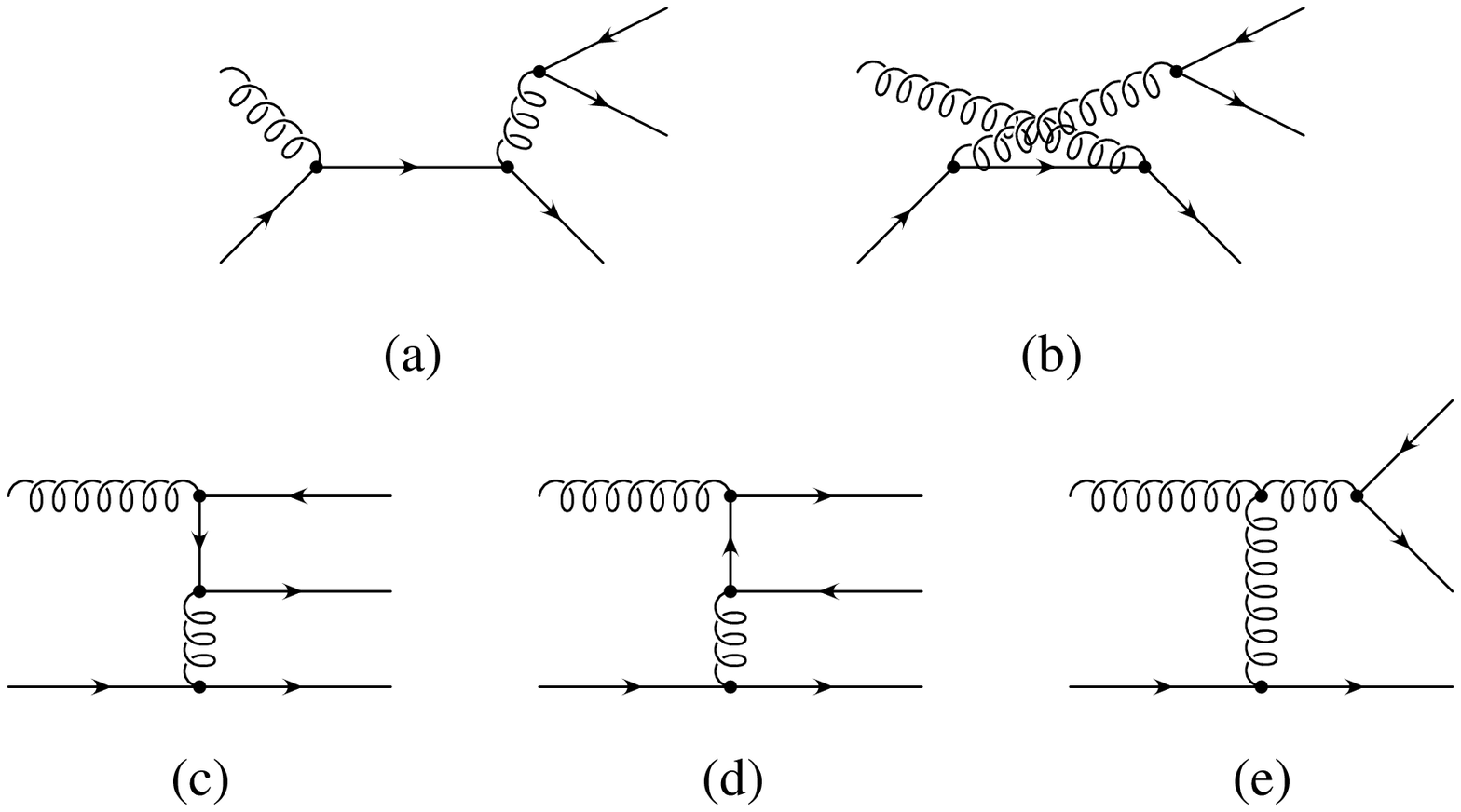}{Origin of the QCD charge
asymmetry in hadroproduction of heavy quarks
through flavor excitation.}
{fig:qg}
%%%%%%%%%%%%%%%

Let us briefly discuss a few important aspect of this 
calculation.  The box amplitude for $q \bar{q} \to Q \bar{Q}$
is ultraviolet finite and the asymmetric contribution to the cross section
of order $\alpha_s^3$ is therefore not affected 
by renormalization, an obvious consequence of the symmetry 
of the lowest order reaction.
The same line of reasoning explains the absence of initial 
state collinear singularities in the limit $m_q \rightarrow 0$
which would have to be absorbed into the (symmetric) lowest 
order cross section.
Infrared singularities require a more careful treatment. 
They are absent in the asymmetric piece of the process in \eq{eq:gq}.
However, real and virtual radiation (Fig.~\ref{fig:qqbar}), if 
considered separately, exhibit infrared divergences, which 
compensate in the sum, corresponding to the inclusive 
production cross section.

The charge asymmetry in the partonic reactions~(\ref{eq:qqbar})
and~(\ref{eq:gq}) implies for example a forward-backward asymmetry
of heavy flavor production in proton-antiproton collisions.
In particular, it leads to a sizeable forward-backward
asymmetry for top production which is dominated by
reaction~(\ref{eq:qqbar}), and can, furthermore, be
scrutinized by studying $t \bar{t}$ production at fixed 
longitudinal momenta and at various partonic energies $\hat{s}$.
However, the charge asymmetry can also be observed in proton-proton
collisions at high energies. In this case one has to 
reconstruct the $t \bar{t}$ restframe and select kinematic regions,
which are dominated by $q \bar{q}$ annihilation 
or flavor excitation $g q \to t \bar{t} X$.
Alternatively, one may also study the difference 
in the one-particle inclusive rapidity distribution 
of top versus antitop, which again integrates to zero.

The analysis of these effects allows to improve our 
understanding of the QCD production mechanism. At the same time 
it is important for the analysis of single top production 
through $W b$ fusion. This reaction is charge asymmetric 
as a consequence of weak interactions. Although the final 
states in single top production and hadronic $t \bar{t}$ production 
are different and should in principle be distinguishable,
it is nevertheless mandatory to control the charge asymmetry 
from both sources.

The presence of charge asymmetric contributions in the 
flavor excitation reaction has also been noticed
in~\cite{Ellis:1986ba,Nason:1989zy,Beenakker:1991ma} for $b$
quark production, however without any quantitative statement. 
The charge asymmetry was also investigated in~\cite{Halzen:1987xd}.
In this work only real gluon emission was considered. To 
arrive at a finite result, a gluon energy infrared cut, $E_{cut}$
had to be introduced which leads to arbitrary large results 
with a pronounced dependence on $E_{cut}$ and a different sign 
of the asymmetry compared to our inclusive calculation.

The outline of this paper is as follows. The technical aspects 
of the calculations will be presented in section~\ref{sec:tec}.
The asymmetric pieces of real and virtual corrections will be
given, together with the numerical evaluation, the compensation
of infrared singularities from real and virtual radiation, and
the results at the partonic level.
The implications for hadronic collisions, proton-proton as well 
as proton-antiproton, will be studied in section~\ref{sec:hadronic}.
A brief summary and our conclusions will be given in
section~\ref{sec:conclusions}.

The main thrust of this paper is toward the study of top 
quarks. Nevertheless all the results are in principle applicable to 
bottom and charm quarks. In practice, of course kinematical 
regions have to be selected, which are dominated by $q \bar{q}$
annihilation or flavor excitation. For $b$ quarks, furthermore,
the dilution effects of mixing must be included. 
We will comment on these points in more detail below.

\section{Amplitudes and partonic cross section}
\label{sec:tec}

As we shall see below, the dominant contribution to the charge 
asymmetry originates from $q \bar{q}$ annihilation, namely from the 
asymmetric piece in the interference between the Born 
amplitude for $q \bar{q} \to Q \bar{Q}$ (Fig.~\ref{fig:qqbar}d) and the 
one loop corrections to this reaction (Fig.~\ref{fig:qqbar}c),
which must be combined with the interference term between
initial state and final state radiation
(Fig.~\ref{fig:qqbar}a,\ref{fig:qqbar}b).
The corresponding contribution to the rate is conveniently 
expressed by the absorptive contributions (cuts) of the 
diagrams depicted in Fig.~\ref{fig:cut}a-\ref{fig:cut}e.

%%%%%%%%%%%%%%%
\mafigura{12 cm}{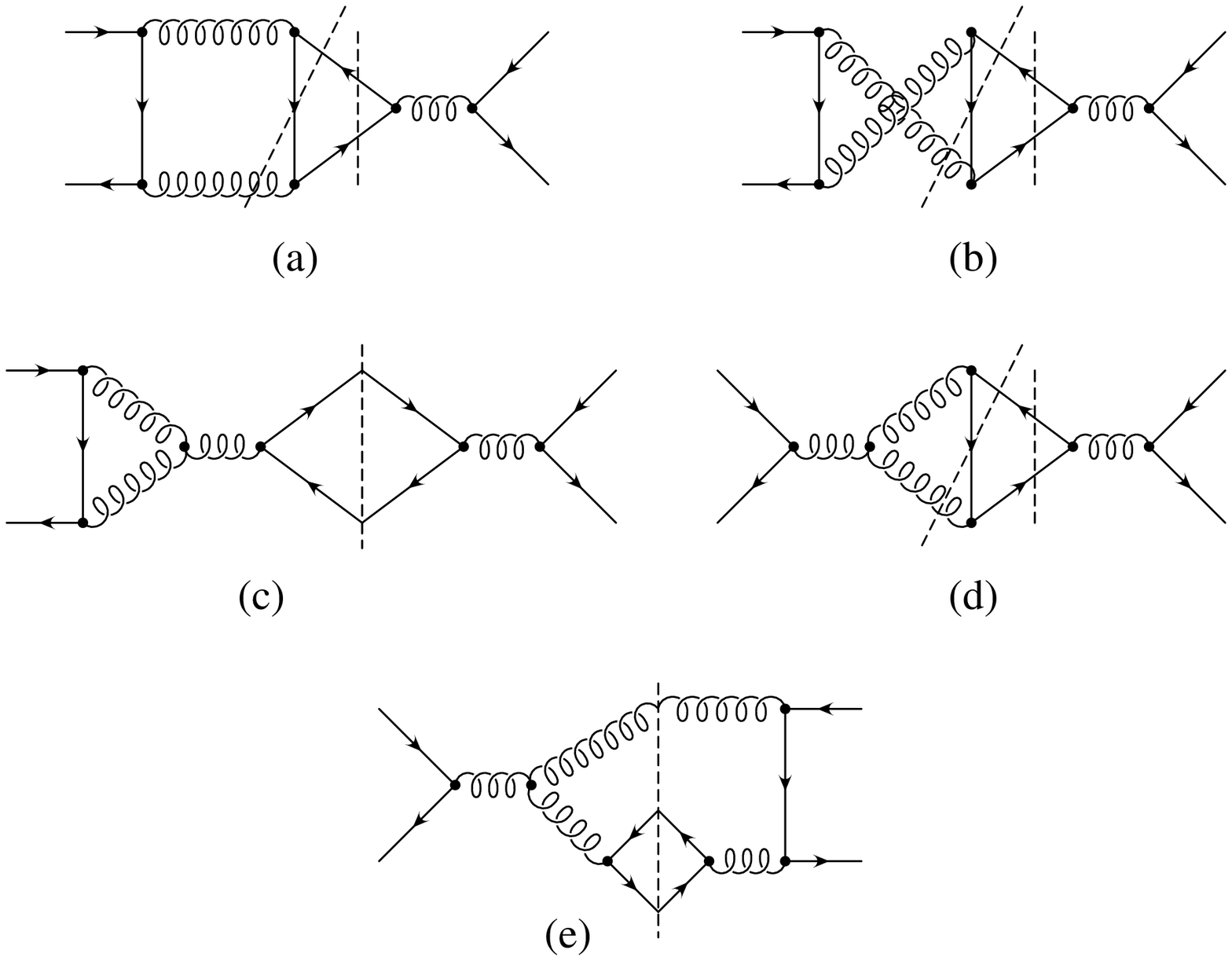}{Cut diagrams.}
{fig:cut}
%%%%%%%%%%%%%%%

However, only Fig.~\ref{fig:cut}a plus the crossed Fig.~\ref{fig:cut}b
are relevant for the charge asymmetric piece.
Fig.~\ref{fig:cut}c, \ref{fig:cut}d and \ref{fig:cut}e, on the other hand 
lead to a symmetric contribution only. 
This can be seen as follows: the color factors corresponding 
to Fig.~\ref{fig:cut}a and~\ref{fig:cut}b (after averaging over 
initial and summing over final states) respectively are given by 
\bea
{\cal C}_{3a} &=& \frac{1}{N_C^2} 
Tr \left( \frac{\lambda^a}{2} \frac{\lambda^b}{2} \frac{\lambda^c}{2} \right)
Tr \left( \frac{\lambda^a}{2} \frac{\lambda^c}{2} \frac{\lambda^b}{2} \right)
= \frac{1}{16 N_C^2} (f_{abc}^2+d_{abc}^2)~, \non \\
{\cal C}_{3b} &=& \frac{1}{N_C^2} 
Tr \left( \frac{\lambda^a}{2} \frac{\lambda^b}{2} \frac{\lambda^c}{2} \right)
Tr \left( \frac{\lambda^b}{2} \frac{\lambda^c}{2} \frac{\lambda^a}{2} \right)
= \frac{1}{16 N_C^2} (-f_{abc}^2+d_{abc}^2)~,
\eea
where $N_C=3$, $f_{abc}^2=24$ and $d_{abc}^2=40/3$.
Without color factors the contributions to 
the cross section from Fig.~\ref{fig:cut}a and~\ref{fig:cut}b
are related by 
\beq
d\sigma_{3a}(Q,\bar{Q}) = - d\sigma_{3b}(\bar{Q},Q)~,
\eeq
which holds true both for two and three 
particle cuts. The asymmetric piece thus originates from the $d_{abc}^2$
term, and its form is thus equivalent to the corresponding 
QED reaction with the replacement of the 
quark charges and QED coupling by the 
color factor 
\beq
\alpha_{QED}^3 Q_q^3 Q_Q^3 \rightarrow \frac{1}{N_C^2}\frac{1}{16} 
d_{abc}^2 \alpha_s^3~.
\eeq
The production cross section, on the other hand is obtained from 
the corresponding QED process through the replacement
\beq
\alpha_{QED}^2 Q_q^2 Q_Q^2 \rightarrow \frac{1}{N_C^2} N_C T_F C_F \alpha_s^2~,
\eeq
with $T_F=1/2$, $C_F=4/3$.
The QCD asymmetry is thus obtained from the QED results by the 
replacement
\beq
\alpha_{QED} Q_q Q_Q \rightarrow \frac{d_{abc}^2}{16 N_C T_F C_F}
\alpha_s = \frac{5}{12} \alpha_s~.
\eeq

Let us note in passing that the cuts through diagrams involving the 
triple gluon coupling 
(Fig.~\ref{fig:cut}c, \ref{fig:cut}d, \ref{fig:cut}e) lead to 
charge symmetric terms.
For Fig.~\ref{fig:cut}c and \ref{fig:cut}e this can be seen as follows:
its contribution involves the factor 
$Tr\{\gamma^{\alpha}(Q\!\!\!\!/ \: + M) \gamma^{\beta} 
(\bar{Q}\!\!\!\!/ \: - M) \}$
which is evidently symmetric under the exchange  of $Q$ and $\bar{Q}$.
The remainder of the diagram depends on $Q+\bar{Q}$ only, which 
leads to charge symmetry of the whole amplitude.
The same line of reasoning applies to Fig.~\ref{fig:cut}d,
as far as the exchange of the initial quarks are concerned.
Charge conjugation invariance then implies, that also 
Fig.~\ref{fig:cut}d is symmetric under the exchange between $Q$ 
and $\bar{Q}$. These terms have to be combined with the charge 
symmetric $C_F C_A$ terms from Fig.~\ref{fig:cut}a and \ref{fig:cut}b
to yield a gauge invariant combination.

Although the relevant ingredients for the charge asymmetric piece are 
already listed in the original QED publications
\cite{Berends:1973,Berends:1983dy}
(and the later works on hadronic heavy flavor production
in~\cite{Ellis:1986ba,Nason:1989zy,Beenakker:1991ma,Halzen:1987xd})
the compact formulae shall be listed in the appendix 
for completeness and convenience of the reader.
In a first step virtual and soft radiation are combined, with
a cut on the gluon energy, $E^g_{cut}$ (see appendix, \eq{eq:softvirtual}).
The logarithmic divergence of this result for small $E^g_{cut}$
is cancelled by the corresponding divergence for real radiation.
A particularly compact formula for the asymmetric piece of the 
hard radiation is given in~\eq{eq:hard}. To obtain finally
the asymmetric piece of the inclusive cross section for 
\beq
q + \bar{q} \to Q + X~,
\eeq
the integral over the real gluon spectrum is performed 
numerically.

The differential charge asymmetry at the partonic level can 
then be defined through 
\beq
\hat{A}(\cos \hat{\theta}) = 
\frac{N_t(\cos \hat{\theta})-N_{\bar{t}}(\cos \hat{\theta})}
     {N_t(\cos \hat{\theta})+N_{\bar{t}}(\cos \hat{\theta})}~,
\eeq
where $\hat{\theta}$ denotes the top quark production angle in 
the $q \bar{q}$ restframe and $N(\cos \hat{\theta}) =
d\sigma/d\Omega (\cos \hat{\theta})$.
Since $N_{\bar{t}}(\cos \hat{\theta}) = N_t(-\cos \hat{\theta})$
as a consequence of charge conjugation symmetry, 
$\hat{A}(\cos \hat{\theta})$ can also be interpreted as
a forward-backward asymmetry of top quarks.
In this case the denominator is of course given by the 
Born cross section for the reaction 
$q \bar{q} \to Q \bar{Q}$, (see \eq{eq:bornqq}).
In Fig.~\ref{fig:sfix}, $\hat{A}(\cos \hat{\theta})$
is displayed for $\sqrt{\hat{s}}=400$~GeV, $600$~GeV and $1$~TeV
for $m_t = 175$~GeV.
For completeness we also display the result for $b \bar{b}$
production at $\sqrt{\hat{s}}=400$~GeV
with $m_b=4.6$~GeV.
The strong coupling constant is evaluated at the scale
$\mu=\sqrt{\hat{s}}/2$ from $\alpha_s(m_Z) = 0.118$.

%%%%%%%%%%%%%%%
\mafigura{8 cm}{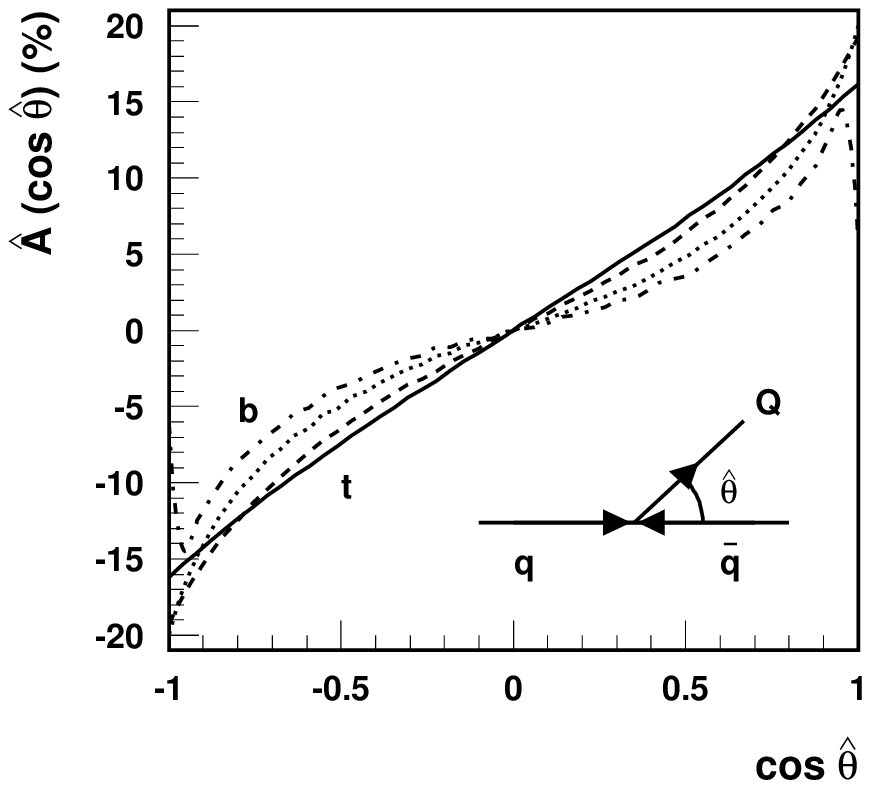}{Differential charge asymmetry 
in top quark pair production for fixed partonic center of
mass energies $\sqrt{\hat{s}}=400$ GeV (solid),
$600$ GeV (dashed) and $1$ TeV (dotted).
We also plot the differential asymmetry for b-quarks 
with $\sqrt{\hat{s}}=400$ GeV (dashed-dotted).}
{fig:sfix}
%%%%%%%%%%%%%%%
%%%%%%%%%%%%%%%
\mafigura{8 cm}{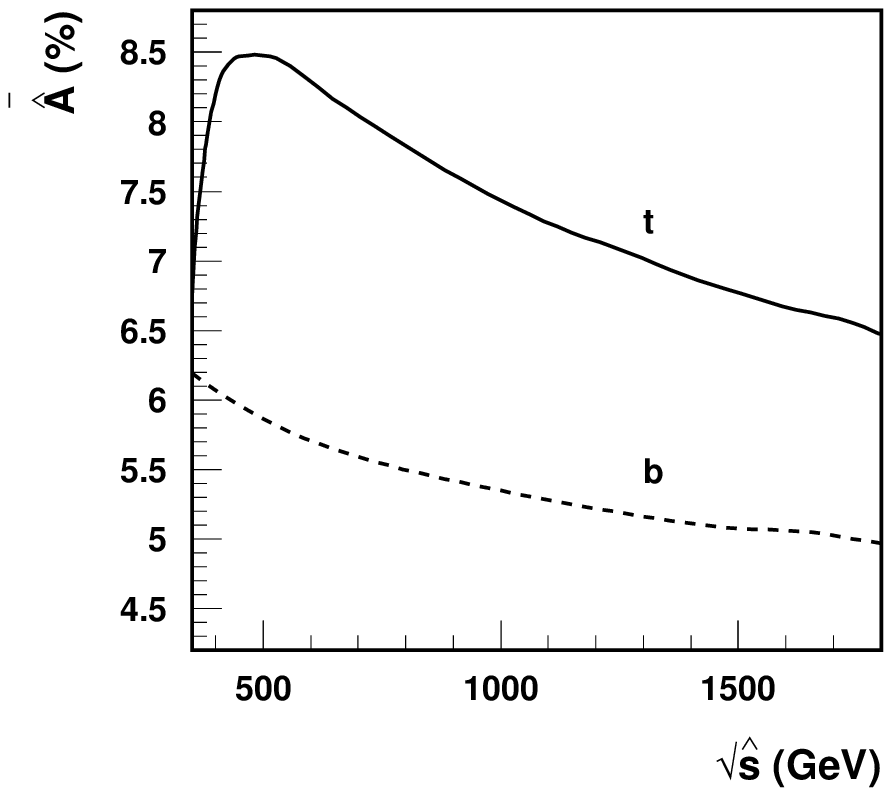}{Integrated charge
asymmetry as a function of the partonic center of mass energy
for top and bottom quark pair production.}
{fig:sdistr}
%%%%%%%%%%%%%%%

The integrated charge asymmetry
\beq
\bar{\hat{A}} = 
\frac{N_t(\cos \hat{\theta} \geq 0)
    - N_{\bar{t}}(\cos \hat{\theta} \geq 0)}
     {N_t(\cos \hat{\theta} \geq 0)
    + N_{\bar{t}}(\cos \hat{\theta} \geq 0)}~,
\eeq
is shown in Fig.~\ref{fig:sdistr} as a function of $\sqrt{\hat{s}}$.
With a typical value around $6-8.5 \%$  it should be well 
accessible in the next run of the TEVATRON.

As mentioned already in the introduction, the asymmetric piece does not 
exhibit a divergence, even in the limit of vanishing initial 
quark mass; in other words, no collinear singularities arise.
The virtual plus soft radiation on one hand and the real 
hard radiation on the other contribute 
with opposite signs, with the former always larger than 
the later which explains the difference in sign 
between our result and~\cite{Halzen:1987xd}.

%%%%%%%%%%%%%%%
\mafigura{8 cm}{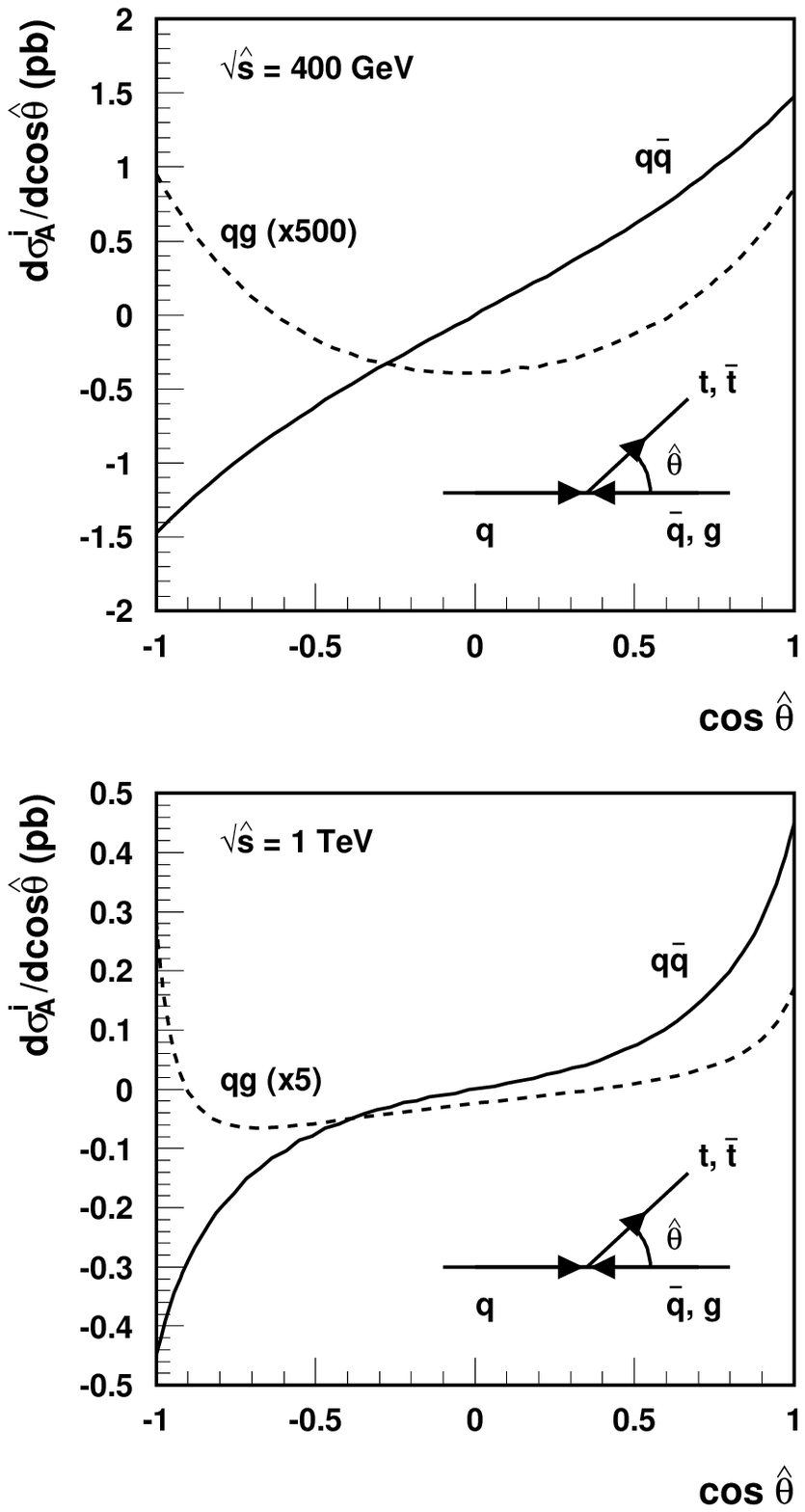}{Asymmetric parts of the differential
top quark pair production cross section
from $q\bar{q}$ and $qg$ initiated processes
for fixed partonic center of
mass energies $\sqrt{\hat{s}}=400$~GeV and $1$~TeV.}
{fig:sfixy}
%%%%%%%%%%%%%%%

Before moving to the application of these results by folding with the 
parton distribution functions let us first discuss the charge 
asymmetry in the quark-gluon induced reaction in~\eq{eq:gq}.
The cross section for this reaction is obtained from the amplitudes
depicted in Fig.~\ref{fig:qg}. In fact its antisymmetric piece can be 
obtained by crossing directly from the reaction
$q\bar{q} \to Q\bar{Q}g$ and is given by~\eq{eq:qg}.
Again only the QED like piece contributes to the 
asymmetry, in contrast to those amplitudes induced by the triple 
gluon coupling.
The inclusive cross section for quark production in quark-gluon 
collisions exhibits a collinear divergence, the charge 
asymmetric piece is finite. 
The difference between $Q$ and $\bar{Q}$ production (for 
fixed initial $q$ and gluon directions) should not be 
confused with an asymmetry in the angular distribution of $Q$
(or $\bar{Q}$) , which is a trivial consequence of the asymmetric 
initial state configuration.

The charge asymmetric pieces as a function of the 
scattering angle
\beq
\frac{1}{2} \left(
\frac{d\sigma(q \bar{q} \to Q X)}{d\cos \hat{\theta}} 
- \frac{d\sigma(q \bar{q} \to \bar{Q} X)}{d\cos \hat{\theta}} 
\right) \equiv \frac{d\sigma^{q\bar{q}}_A}{d\cos \hat{\theta}}~,
\eeq
\beq
\frac{1}{2} \left(
\frac{d\sigma(qg \to Q X)}{d\cos \hat{\theta}} 
- \frac{d\sigma(qg \to \bar{Q}X)}{d\cos \hat{\theta}} 
\right) \equiv \frac{d\sigma^{qg}_A}{d\cos \hat{\theta}}~,
\eeq
are shown in Fig.~\ref{fig:sfixy} for a variety of partonic 
energies $\sqrt{\hat{s}}$ with $m_t=175$~GeV.

%%%%%%%%%%%%%%%
\mafigura{8 cm}{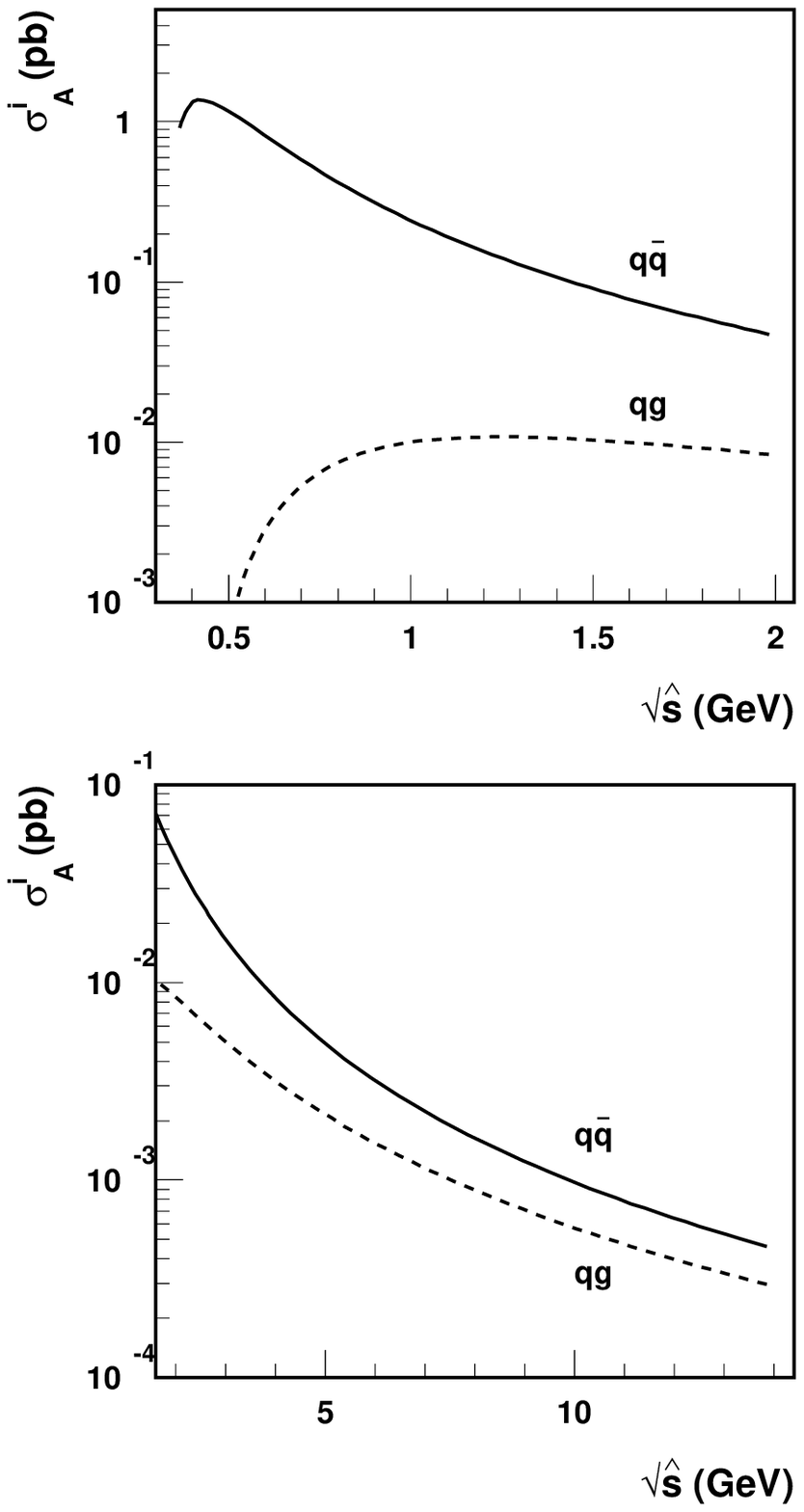}{Integrated
charge asymmetric parts of the top quark pair production cross section
from $q\bar{q}$ and $qg$ initiated processes
as a function of the partonic center of mass energy.}
{fig:sdistry}
%%%%%%%%%%%%%%%

The asymmetric contributions integrated 
in the forward-backward direction
\beq
\sigma_A^i = 
\int_0^1 \frac{d\sigma^i_A}{d\cos \hat{\theta}} d\cos \hat{\theta} 
- \int_{-1}^0 \frac{d\sigma^i_A}{d\cos \hat{\theta}} d\cos \hat{\theta}~, 
\qquad i=q\bar{q},qg~,
\eeq
are shown in Fig.~\ref{fig:sdistry}.
Clearly, at the TEVATRON energies with $\sqrt{\hat{s}} < 2$~TeV,
the dominant asymmetric contribution comes from the $q \bar{q}$
initiated processes, even more so, since the quark-gluon luminosity
in the relevant kinematic region is far below the one for
quark-antiquark reactions. 
Furthermore, the difference between $Q$ and $\bar{Q}$ 
production in quark-gluon collision 
does not exhibit a marked forward-backward asymmetry,
which suppresses the $q g$ induced asymmetry even further.

%%%%%%%%%%%%%%%
\mafigura{12 cm}{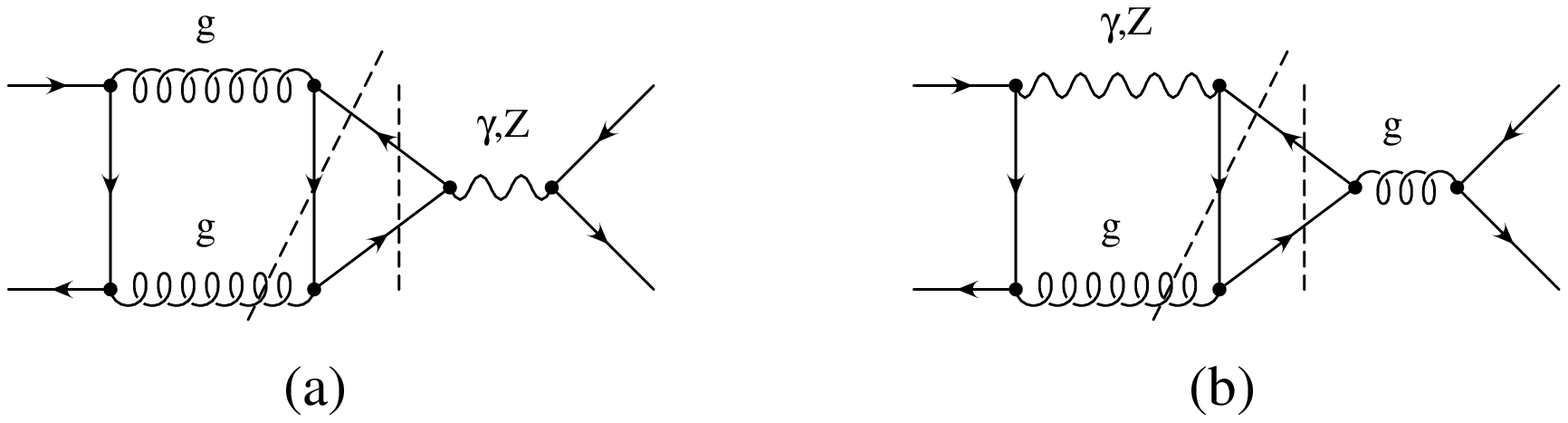}
{Representative diagrams contributing to the QCD neutral current
interference term.}
{fig:gammaZ}
%%%%%%%%%%%%%%%

In addition to the pure QCD amplitudes also a mixed QCD-electroweak 
interference term will lead to an asymmetric contribution
to the $q\bar{q}$ process. 
The QCD box diagram, Fig.~\ref{fig:qqbar}c, can also give 
rise to $t \bar{t}$ in a color singlet configuration, 
which in turn interferes with $t \bar{t}$ production through the 
photon or $Z$ (Fig.~\ref{fig:gammaZ}a).
A similar consideration applies to interference between 
initial and final state radiation.
The resulting asymmetry is obtained from the QCD asymmetry through the 
following replacement 
\beq
\frac{\alpha_s}{2} \left( \frac{d_{abc}}{4} \right)^2
\rightarrow 
\alpha_{QED} \left( Q_t Q_q + 
\frac{(1-\frac{\displaystyle 8}{\displaystyle 3} s^2_W)(2 I_q-4 Q_q s^2_W)}
{16 s^2_W c^2_W}  
\frac{1}{1-\frac{\displaystyle m_Z^2}{\displaystyle \hat{s}}} \right)~.
\label{eq:gammaZ} 
\eeq
Another QED-electroweak term originates from the interference 
between the gluon-$\gamma$ box and gluon-$Z$ box 
respectively with the QCD Born amplitude (Fig.~\ref{fig:gammaZ}b). 
The result for this piece~\footnote{This small term had been 
neglected in~\cite{Kuhn:1998jr}.} is also given by~\eq{eq:gammaZ}.
In total this leads to an increase of the asymmetry as given by
pure QCD by a factor $1.09$. This change is thus smaller than 
uncalculated higher order corrections.

\section{Hadronic collisions}
\label{sec:hadronic}

The asymmetry can in principle be studied experimentally in 
the partonic restframe, as a function of $\hat{s}$, by 
measuring the invariant mass of the $t \bar{t}$ system 
plus an eventually radiated gluon. 
It is, however, also instructive
to study the asymmetry in the laboratory frame by folding 
the angular distribution with the structure
functions~\cite{Martin:1996as,Lai:1997mg}.
For proton-antiproton collisions it is convenient to consider 
the forward-backward asymmetry as function of the 
production angle in the center of mass system.
The differential asymmetry for $\sqrt{s}=2$~TeV
is shown in Fig.~\ref{fig:PDFdistr}
which displays separately the contribution from 
$q\bar{q}$ and $qg$ (plus $\bar{q}g$) initiated reactions.
The denominator includes both $q\bar{q}$ and $gg$ initiated
processes in lowest order.
The numerator is evidently dominated by quark-antiquark annihilation 
as anticipated in~\cite{Kuhn:1998jr}
as can be seen from Fig.~\ref{fig:PDFdistr}.
Inclusion of mixed QCD/electroweak interference term enhances 
the prediction by a factor $1.09$.

At this point we have to emphasize that both numerator and
denominator are evaluated in leading order (LO).
The next-to-leading (NLO) corrections to the $t\bar{t}$
production cross section are known to be large~\cite{Bonciani:1998vc},
around $30\%$ or even more.
In the absence of NLO corrections for the numerator we nevertheless
stay with the LO approximation in both numerator and denominator,
expecting the dominant corrections from collinear emission to cancel.
However, from a more conservative point of view an uncertainty of 
around $30\%$ has to be assigned to the prediction for the asymmetry.

For the total charge asymmetry  at $\sqrt{s}=1.8$~TeV we predict 
\beq
\bar{A} = 
\frac{N_t(\cos \theta \geq 0)
    - N_{\bar{t}}(\cos \theta \geq 0)}
     {N_t(\cos \theta \geq 0)
    + N_{\bar{t}}(\cos \theta \geq 0)} = 4.8 - 5.8 \%~,
\eeq
where different choices of the structure function and different
choices of the factorization and renormalization scale,
$\mu = m_t/2$ and $\mu = 2 m_t$,
have been considered and the factor $1.09$ is included.
An increase in the center of mass energy to $2$~TeV leads to a slight 
decrease of our prediction to $4.6-5.5 \%$.

%%%%%%%%%%%%%%%
\mafigura{8 cm}{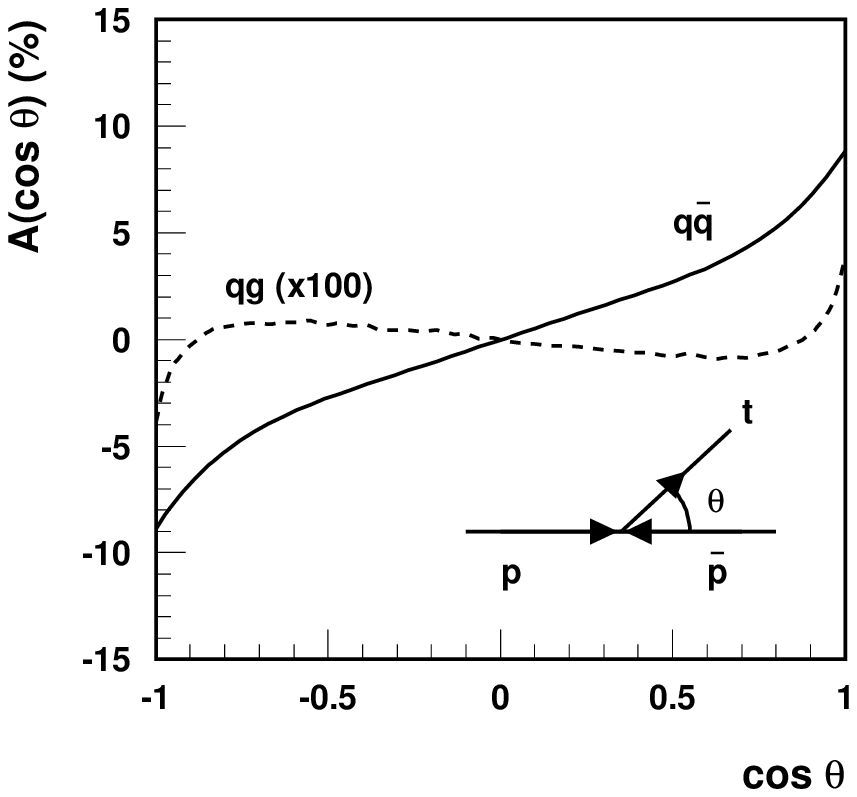}
{Differential charge asymmetry in the proton-antiproton restframe,
$\sqrt{s}=2$~TeV, using the CTEQ-1 structure function, $\mu=m_t$. 
The contributions from $q\bar{q}$ and $qg$ (plus $\bar{q}g$) 
initiated processes are shown separately.}
{fig:PDFdistr}
%%%%%%%%%%%%%%%

For illustrative purpose in Fig.~\ref{fig:TEV_CTEQ1_cm} the
$q\bar{q}$ and $qg (\bar{q}g)$ induced
contributions in the partonic restframe are also displayed separately
in the $x=x_1-x_2$ and $\hat{s}$ plane for proton-antiproton collisions
with $\mu=\sqrt{\hat{s}}/2$.
Furthermore, as characteristical example the relative amount 
of gluon fusion as function of $x=x_1-x_2$ and $\hat{s}$
is shown in the two dimensional distribution of Fig.~\ref{fig:relamount}.
In regions of larger $q\bar{q}$ and correspondingly smaller $gg$
induced reactions a larger asymmetry is expected.

Bottom quark production at the LHC or TEVATRON is of course
dominated by gluon fusion.
The forward-backward asymmetry from $q\bar{q}$ and $qg(\bar{q}g)$
reactions is thus negligible, at least as far as the total 
cross section is concerned.
However, the $b\bar{b}(g)$ final state with $\hat{s}$
sufficiently large, say above $300$~GeV is again dominated by 
$q\bar{q}$ annihilation and a sizeable asymmetry is 
predicted in this kinematical region.
Selecting for example $\sqrt{\hat{s}} \geq 300$~GeV
and $\mid \cos \theta \mid < 0.9$ one predicts ($\sqrt{s}=2$~TeV)
\beq
\bar{A} = 4.3 - 5.1 \%~,
\eeq
which should be accessible by experiment.
A factor $0.96$, from the QCD/electroweak interference,
has been also included.

Top-antitop production in proton-proton collisions at the 
LHC is, as a consequence of charge conjugation symmetry,
forward-backward symmetric if the laboratory frame 
is chosen as the reference system. However, by selecting
the invariant mass of the $t \bar{t} (+g)$ system and its
longitudinal momentum appropriately, one can easily constrain
the parton momenta such that a preferred direction is 
generated for quark-antiquark reactions.

%For some of the more extreme kinematic regions a sizeable
%difference between top and antitop production can be 
%observed at the LHC (Fig.~\ref{fig:LHC_CTEQ1_cm}).
%The production cross section {\it per se}, which is decisive 
%for the possibility of measuring the asymmetry in 
%these regions is displayed in Fig.~\ref{}.
%In Fig.~\ref{fig:LHC_CTEQ1_cm} the contribution from $q\bar{q}$
%and $qg$ induced reactions are displayed separately. 
%In practice, only the region with $\hat{s}$ below $2$~TeV will 
%be observable, in particular at large $x$.

For some of the more extreme kinematic regions,
namely large $x$ and/or large $\hat{s}$, a sizeable difference
between top and antitop production can be observed at the LHC.
In Fig.~\ref{fig:LHC_CTEQ1_cm} the contributions from
$q \bar{q}$ and $q g$ ($\bar{q}g$ induced reactions are displayed  separately.
The production cross section {\it per se}, which is decisive 
for the possibility of measuring the asymmetry in 
these regions is displayed in Fig.~\ref{fig:singleinclusive}b.
In practice, only the region with $\hat{s}$ below $2$~TeV will 
be observable, in particular at large $x$.

The asymmetry, as displayed in Fig.~\ref{fig:LHC_CTEQ1_cm}
as a function of $x=x_1-x_2$ and $\hat{s}$, is defined
in the $t\bar{t}(g)$ restframe. From this it may 
seem that the reconstruction of both $t$, $\bar{t}$ and even the 
gluon is required for the study of the charge asymmetry in $pp$
collisions. However, also the difference between 
the single particle inclusive distribution of $t$ and $\bar{t}$
respectively may provide evidence for the charge asymmetry. 
This can be easily understood from Fig.~\ref{fig:preferred}.
Production of $t\bar{t}(g)$ with negative $x$ is dominated 
by initial $\bar{q}$ with small $x_1$ and $q$ with large $x_2$.
The charge asymmetry implies that $Q(\bar{Q})$ is 
preferentially emitted into the direction of $q(\bar{q})$,
Fig.~\ref{fig:preferred}a, \ref{fig:preferred}b.
The same line of reasoning is applicable for 
positive $x$, with $Q(\bar{Q})$ again preferentially emitted in 
the direction of $q(\bar{q})$, and the role of $x_1$ and 
$x_2$ reversed. In total this leads to a slight preference 
for centrally  produced antiquarks and quarks slightly  
dominant in the forward and backward direction, 
i.e., at large positive and negative rapidities. 

%%%%%%%%%%%%%%%
\mafigura{8 cm}{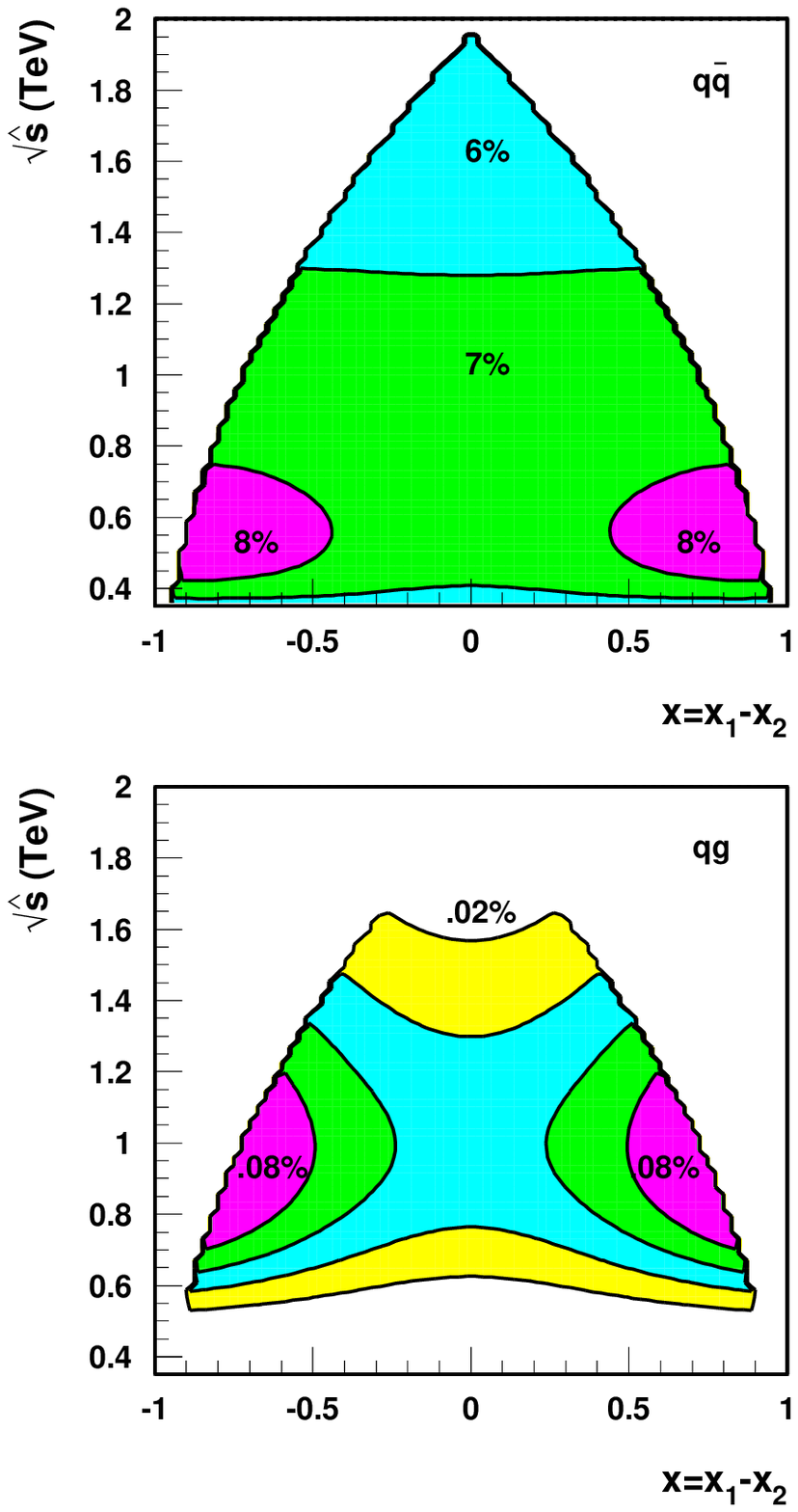}{
Contributions from $q\bar{q}$ and $qg (\bar{q}g)$ induced 
reactions to the charge asymmetry in proton-antiproton collisions,
$\sqrt{s}=2$~TeV, as a function of $x_1-x_2=2 P_3(t \bar{t} g)/\sqrt{s}$
and $\hat{s}$. Partonic restframe (CTEQ-1).}
{fig:TEV_CTEQ1_cm}
%%%%%%%%%%%%%%%
%%%%%%%%%%%%%%%
\mafigura{8 cm}{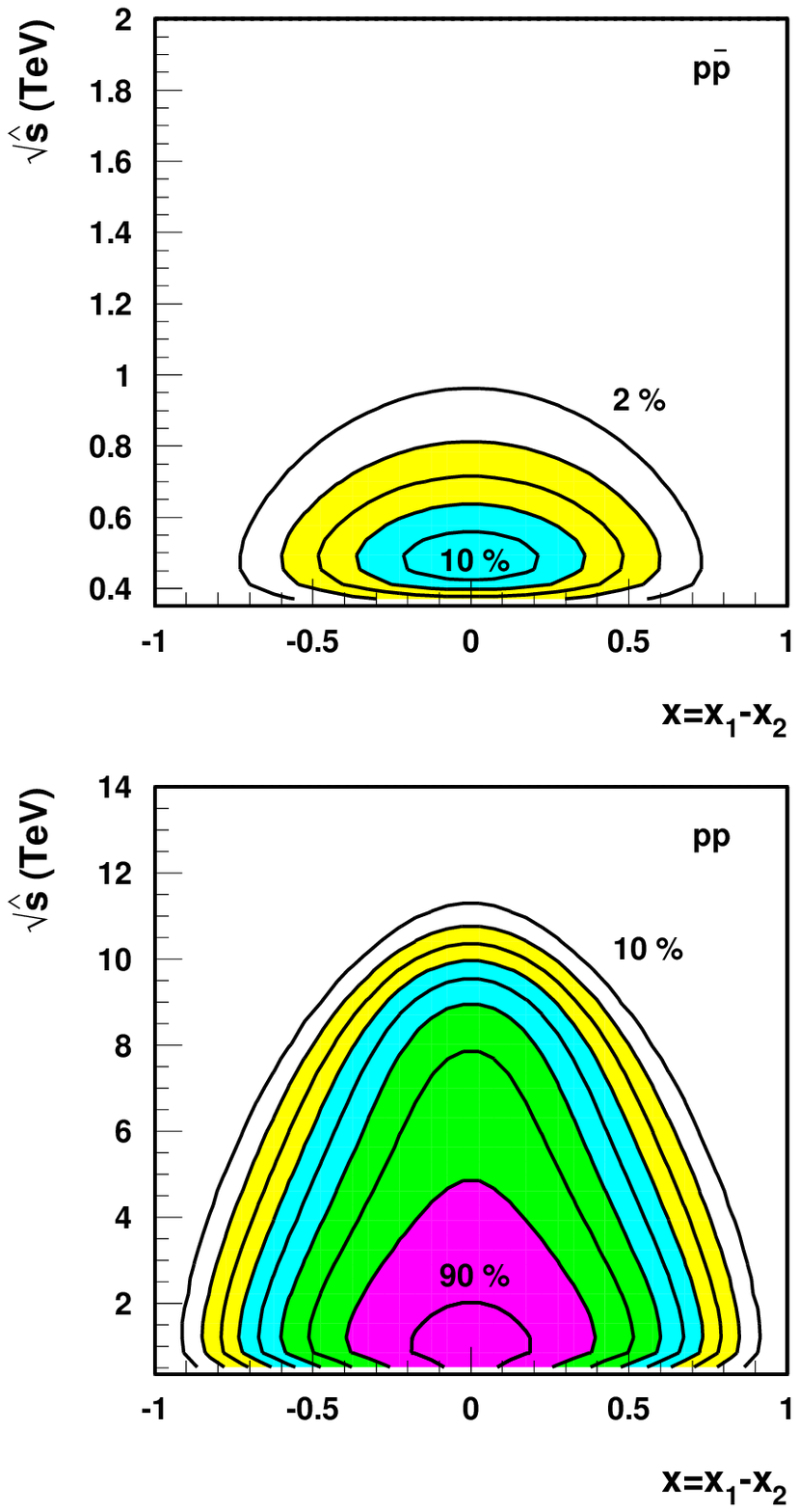}{
Relative amount of gluon-gluon initiated processes as a function of 
$x_1-x_2=2 P_3(t \bar{t} g)/\sqrt{s}$ and $\hat{s}$
in lowest order,
for $\sqrt{s}=2$~TeV in proton-antiproton collisions
and $\sqrt{s}=14$~TeV in proton-proton collisions.
Contour lines go from $2\%$ (white) to $10\%$ (dark) in steeps 
of $2\%$ in the former case and from 
from $10\%$ (white) to $90\%$ (dark) in steeps 
of $10\%$ in the later (CTEQ-1).}
{fig:relamount}
%%%%%%%%%%%%%%%
%%%%%%%%%%%%%%%
\mafigura{8 cm}{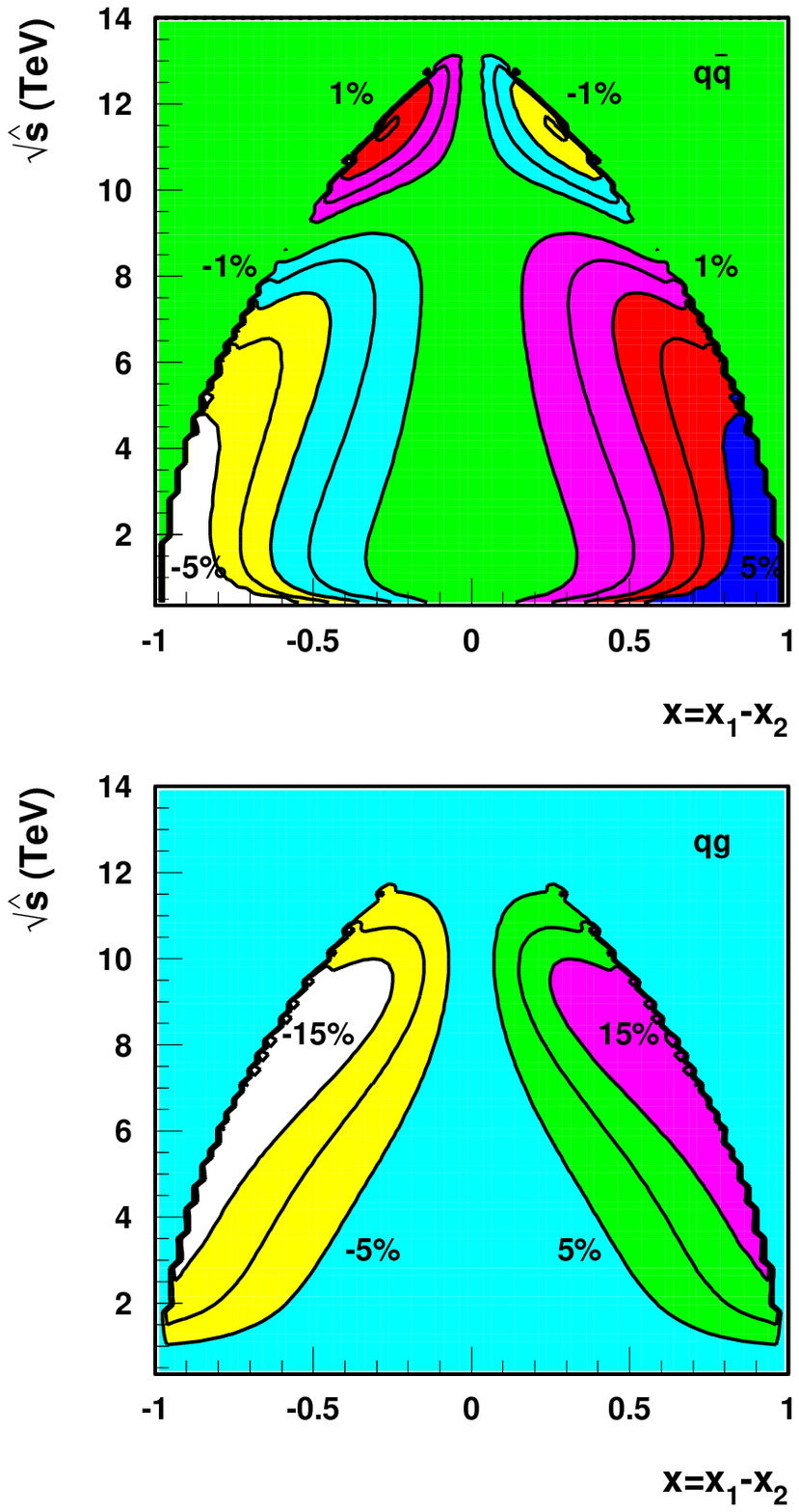}{
Contributions from $q\bar{q}$ and $qg (\bar{q}g)$ induced 
reactions to the charge asymmetry in proton-proton collisions,
$\sqrt{s}=14$~TeV, as a function of $x_1-x_2=2 P_3(t \bar{t} g)/\sqrt{s}$
and $\hat{s}$. Partonic restframe (CTEQ-1).}
{fig:LHC_CTEQ1_cm}
%%%%%%%%%%%%%%%

\pagebreak

%%%%%%%%%%%%%%%
\mafigura{10 cm}{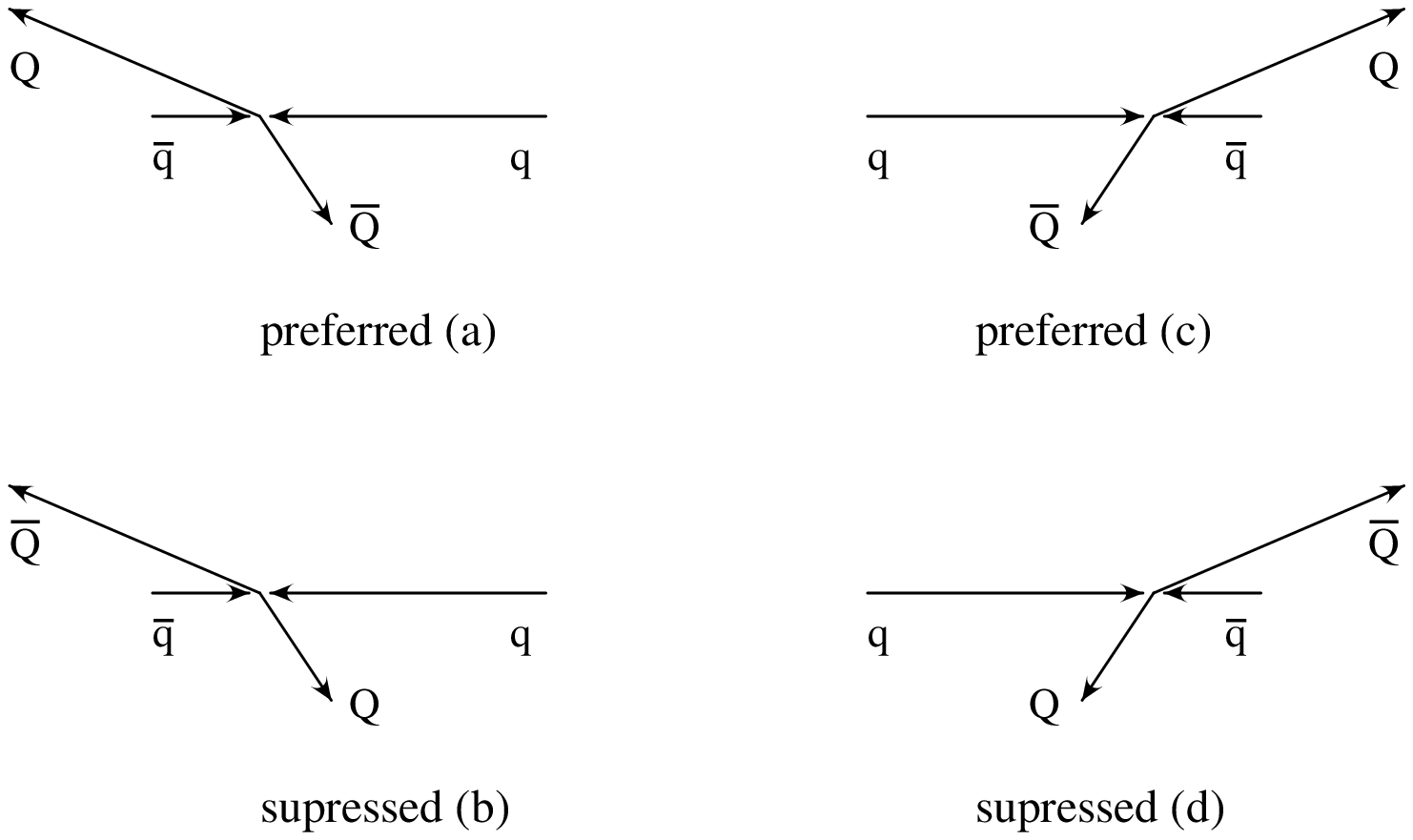}
{Typical configuration of momenta for top and antitop 
production through quark annihilation in the region of large 
parallel momenta.}
{fig:preferred}
%%%%%%%%%%%%%%%

%%%%%%%%%%%%%%%
\mafigura{8 cm}{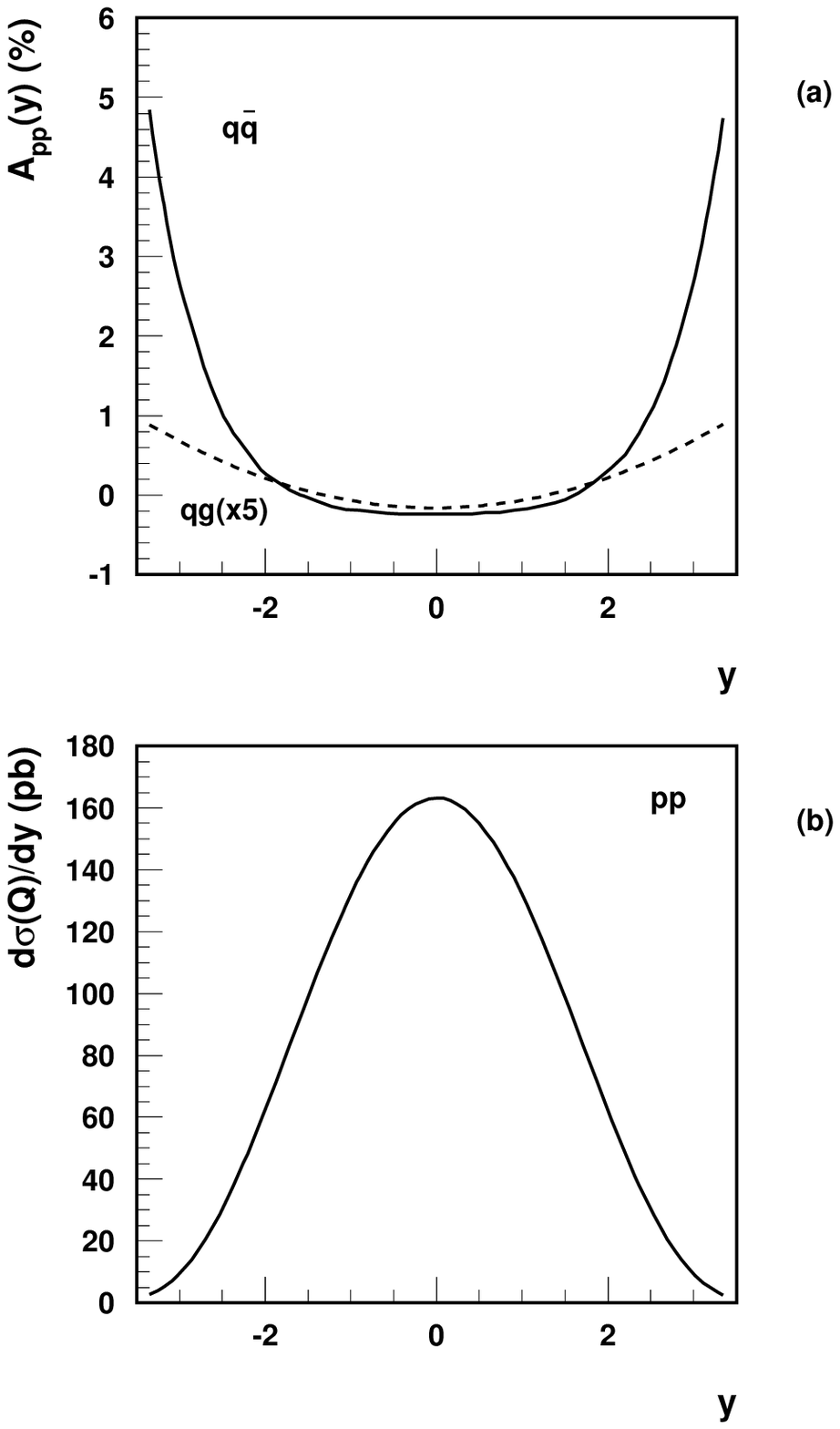}
{Rapidity distribution of charge asymmetry (a) and
total cross section at Born order (b) of top quark 
production in proton-proton collisions, $\sqrt{s}=14$~TeV and
$\mu=m_t$. Contributions from $q\bar{q}$ fusion 
and flavor excitation, $qg$($\bar{q}g$), are shown 
separately. Laboratory frame (CTEQ-1).}
{fig:singleinclusive}
%%%%%%%%%%%%%%%

\pagebreak

The differential charge asymmetry 
\beq
A_{pp}(y) = \frac{\frac{\displaystyle dN(Q)}{\displaystyle dy}
- \frac{\displaystyle dN(\bar{Q})}{\displaystyle dy}}
{\frac{\displaystyle dN(Q)}{\displaystyle dy} 
+ \frac{\displaystyle dN(\bar{Q})}{\displaystyle dy}}~,
\eeq
is shown in Fig.~\ref{fig:singleinclusive}a for top quark
production at the LHC ($\sqrt{s}=14$~TeV).
As expected, a sizeable charge asymmetry is predicted 
in the region of large rapidity. It remains to be seen, 
if the low event rates in these extreme regions will permit 
the observation of this effect. The quark-gluon process is 
again negligible.

\section{Conclusions}
\label{sec:conclusions}

The difference between the distributions of quarks versus antiquarks
produced in hadronic collisions has been investigated.
This asymmetry is particularly relevant for the production 
of top quarks in suitably chosen kinematical regions, but can also 
been observed for bottom quarks at large $\hat{s}$. 
At the partonic level it amount up to $10\%$ or even $15\%$.
In proton-antiproton collisions at the TEVATRON the integrated 
forward-backward asymmetry amounts close to $5\%$. 
At the LHC a slight preference for centrally 
produced antitop is predicted, with top quarks more abundant at 
large positive and negative rapidities.

\appendix 
\section{Basic formulae}

For completeness we summarize here the charge antisymmetric 
contributions to the heavy quark production cross section. 
The charge asymmetric piece of the hard gluon radiation process 
\beq
q(p_1)+\bar{q}(p_2) \rightarrow Q(p_3) + \bar{Q}(p_4) + g(p_5)~,
\label{ap:qqbar}
\eeq
defined as
\beq
d\sigma^{q\bar{q}}_A \equiv 
\frac{1}{2} \left(
d\sigma(q \bar{q} \to Q X) - d\sigma(q \bar{q} \to \bar{Q} X)
\right)~,
\eeq
is given by 
\bea
\frac{d\sigma^{q\bar{q},hard}_A}{dy_{35}\: dy_{45}\: d\Omega} &=&
\frac{\alpha_s^3}{4\pi \hat{s}} \:
\frac{d^2_{abc}}{16 N_C^2} \:
\frac{1}{y_{12}\: (y_{34}+2m^2)\: y_{35}} \non \\
&\times &  \left\{
\frac{y_{13}}{y_{15}}\left(y_{13}^2+y_{14}^2+y_{23}^2+y_{24}^2
+2m^2(y_{34}+2m^2+y_{12}) \right) 
+ 4 m^2 \: y_{24} \right\} \non \\ \non \\ 
&-&(1 \leftrightarrow 2) - (3 \leftrightarrow 4)
+ (1 \leftrightarrow 2,3 \leftrightarrow 4 )~, 
\label{eq:hard}
\eea
with $N_C=3$ and $d_{abc}^2=40/3$.
All the quantities are normalized to the partonic center of mass
energy $\hat{s}$,
\beq
y_{ij}=2(p_1\cdot p_j)/\hat{s}~, \qquad m^2 = m_Q^2/\hat{s}~.
\eeq
The asymmetry is explicitly driven by the antisymmetric 
exchange of momenta $(p_i \leftrightarrow p_j)$.

On the other hand, soft radiation from \eq{ap:qqbar}
integrated in phase space up to a cut in the soft gluon 
energy, $E_{cut}^g$, plus the virtual corrections to 
the Born process $q\bar{q} \to Q\bar{Q}$ contribute to the 
asymmetry as 
\bea
\frac{d\sigma^{q\bar{q},virt+soft}_A}{d\cos \hat{\theta}} &=& 
\frac{\alpha_s^3}{2 \hat{s}} \:
\frac{d^2_{abc}}{16 N_C^2} \:
\: \beta \: \bigg\{ B(c) - B(-c) \non \\
&+& (1+c^2+4m^2) \left[ 4 \log\left(\frac{1-c}{1+c}\right) \log(2w)
+ D(c) - D(-c) \right]   \bigg\}~,
\label{eq:softvirtual}
\eea 
with
\beq
\beta = \sqrt{1-4m^2}~, \qquad 
c = \beta \cos \hat{\theta}~, \qquad 
w = E_{cut}^g/\sqrt{\hat{s}}~,
\eeq
and the functions $B(c)$, coming from the box contribution,
and $D(c)$, from soft radiation, defined as 
\bea
B(c) &=& \frac{1-c^2-8m^2}{1-c-2m^2}
\log \left(\frac{1-c}{2}\right)
+ (c+2m^2) \left[2 Li_2\left(1-\frac{2m^2}{1-c}\right) 
-\log^2\left(\frac{1-c}{2}\right) \right] \non \\
&+& \frac{4c}{\beta^2} \frac{2-c^2-7m^2}{(1-2m^2)^2-c^2} \: m^2 \: \log(m^2)
+\frac{c}{2} \log^2(m^2) \non \\
&-& \frac{c}{2\beta^3}(1-8m^2+8m^4)\left[ 
\log^2\left(\frac{1-\beta}{1+\beta}\right) 
+ 4 Li_2\left(-\frac{1-\beta}{1+\beta}\right)
+ \frac{\pi^2}{3}\right] - c \: \frac{\pi^2}{6}~,
\eea
\bea
D(c) &=& 
2 Re \left\{ Li_2\left(\frac{-x}{1-y}\right)
- Li_2\left(\frac{1-x}{1-y}\right)
- Li_2\left(\frac{1+x}{y}\right)
+ Li_2\left(\frac{x}{y}\right) \right\} \non \\
&+& \log^2 \left| \frac{y}{1-y}\right|
- Re \: Li_2(x^2) + \frac{1}{2} \log^2 (x^2)
- \log(x^2) \log(1-x^2)~,
\eea
where 
\beq
x = \frac{1-c}{\sqrt{2(1-c-2m^2)}}~, \qquad
y = \frac{1}{2} \left( 1-\beta+\sqrt{2(1-c-2m^2)} \right)~. 
\eeq 
In the limit $m \rightarrow 0$ these functions
simplify considerably
\bea
B(c) &=&
\left[1+c-c \log\left(\frac{1-c}{2}\right)\right]
\log\left(\frac{1-c}{2}\right)~, \non \\
D(c) &=& \log^2\left(\frac{1-c}{2}\right) - 2 Li_2\left(\frac{1-c}{2}\right)~,
\eea
and become also free of final state collinear divergences
since only integrable divergences appear at $c=1$.

The charge asymmetric contribution of the flavor excitation process
\beq
q(p_1)+g(p_2) \rightarrow Q(p_3) + \bar{Q}(p_4) + q(p_5)~,
\eeq
defined as 
\beq
d\sigma^{qg}_A \equiv 
\frac{1}{2} \left(
d\sigma(qg \to Q X) - d\sigma(qg \to \bar{Q}X)
\right)~,
\eeq
is given by 
\bea
\frac{d\sigma^{qg}_A}{dy_{35}\: dy_{45}\: d\Omega} &=&
\frac{\alpha_s^3}{4\pi \hat{s}} \: 
\frac{d^2_{abc}}{16 N_C^2} \:
\frac{1}{y_{15}\: (y_{34}+2m^2)\: y_{23}} \non \\
&\times & \Bigg\{
\left(\frac{y_{13}}{y_{12}} - \frac{y_{35}}{y_{25}}\right)
\left(y_{13}^2+y_{14}^2+y_{35}^2+y_{45}^2
+2m^2(y_{34}+2m^2-y_{15}) \right) \non \\
& & + 4 m^2 \: (y_{45}+y_{14}) \Bigg\} 
- (3 \leftrightarrow 4)~.
\label{eq:qg}
\eea
It is infrared finite and can be obtained just by crossing of
momenta from \eq{eq:hard}.

For the convenience of the reader we also list 
the Born cross section for $q\bar{q}$ and $gg$ fusion
\beq
\frac{d\sigma^{q\bar{q}\rightarrow Q \bar{Q}}}{d\cos \hat{\theta}} = 
\alpha_s^2 \: \frac{T_F C_F}{N_C} \:  
\frac{\pi \beta}{2 \hat{s}}
(1+c^2+4m^2)~,
\label{eq:bornqq}
\eeq
\beq
\frac{d\sigma^{gg\rightarrow Q \bar{Q}}}{d\cos \hat{\theta}} = 
\alpha_s^2 \: \frac{\pi \beta}{2 \hat{s}}  
\left(\frac{1}{N_C(1-c^2)}-\frac{T_F}{2C_F}\right)
\left(1 + c^2 +8 m^2-\frac{32 m^4}{1-c^2}\right)~,
\eeq
where $T_F=1/2$ and $C_F=4/3$.

\vspace{.5cm}

We would like to acknowledge useful discussions with R.K.~Ellis,
T.~Sj\"ostrand and M.~Seymour.
Work supported by BMBF under Contract 057KA92P 
and DFG under Contract Ku 502/8-1.
G.R. acknowledges a postdoctoral fellowship from INFN
and the TTP Institute at the University of Karlsruhe for 
the kind hospitality during the completion of this work.

%\bibliographystyle{utphys}
%\bibliography{asymetry}

\begin{thebibliography}{10}

\bibitem{Catani:1997rn}
S.~Catani, ``{QCD} at high-energies,'' {{\tt hep-ph/9712442}},
and references therein.

\bibitem{Tipton:1996}
P.~Tipton, ``Experimental top quark physics,'' {\em Proceedings of the ICHEP
  96, Warsaw, Poland} (World Scientific, Singapore, 1996), pg.123.

\bibitem{Marchesini:1991ch}
G.~Marchesini {\em et.~al.} {\em Comput. Phys. Commun.} {\bf 67} (1992) 465.

\bibitem{Sjostrand:1994yb}
T.~Sj{\"o}strand {\em Comput. Phys. Commun.} {\bf 82} (1994) 74.

\bibitem{Ellis:1986ba}
R.~K. Ellis {\em in Strong Interactions and Gauge Theories}, ed. J.~Tran Thanh
  Van (Editions Fronti\`ere, Gif-sur-Yvette,1986), pg.339.

\bibitem{Nason:1989zy}
P.~Nason, S.~Dawson, and R.~K. Ellis {\em Nucl. Phys.} {\bf B327} (1989) 49.

\bibitem{Beenakker:1991ma}
W.~Beenakker, W.~L. van Neerven, R.~Meng, G.~A. Schuler, and J.~Smith {\em
  Nucl. Phys.} {\bf B351} (1991) 507.

\bibitem{Halzen:1987xd}
F.~Halzen, P.~Hoyer, and C.~S. Kim {\em Phys. Lett.} {\bf 195B} (1987) 74.

\bibitem{Berends:1973}
F.~A. Berends, K.~J.~F. Gaemers, and R.~Gastmans {\em Nucl. Phys.} {\bf B63}
  (1973) 381.

\bibitem{Berends:1983dy}
F.~A. Berends, R.~Kleiss, S.~Jadach, and Z.~Was {\em Acta Phys. Polon.} {\bf
  B14} (1983) 413.

\bibitem{Martin:1996as}
A.~D. Martin, R.~G. Roberts, and W.~J. Stirling {\em Phys. Lett.} {\bf B387}
  (1996) 419, {{\tt hep-ph/9606345}}.

\bibitem{Lai:1997mg}
H.~L. Lai {\em et.~al.} {\em Phys. Rev.} {\bf D55} (1997) 1280,
{\tt hep-ph/9606399}.

\bibitem{Kuhn:1998jr}
J.~H. K{\"u}hn and G.~Rodrigo {\em Phys. Rev. Lett.} {\bf 81}
  (1998) 49.

\bibitem{Bonciani:1998vc}
R.~Bonciani, S.~Catani, M.~L.~Mangano and P.~Nason,
``NLL resummation of the heavy quark hadroproduction cross-section'', 
{{\tt hep-ph/9801375}},



\end{thebibliography}

\end{document}